\documentclass[%
reprint,
superscriptaddress,
showpacs,preprintnumbers,
 amsmath,amssymb,
 aps,
 pra,
]{revtex4-1}
\usepackage{graphicx}
\usepackage{bm}
\usepackage{booktabs} 
\usepackage{hyperref}

\graphicspath{{./figs/}}
\DeclareMathOperator*{\argmax}{arg\,max}

\begin{document}

\preprint{APS/123-QED}

\title{Investigation of Voronoi diagram based Direction Choices Using Uni- and Bi-directional Trajectory Data}

\author{Yao Xiao}
\affiliation{
School of Traffic and Transportation, Beijing Jiaotong University, 100044 Beijing, China
}%
\affiliation{
Institute for Advanced Simulation, Forschungszentrum J\"ulich, 52425 J\"ulich, Germany
}%
\author{Mohcine Chraibi}
\affiliation{
Institute for Advanced Simulation, Forschungszentrum J\"ulich, 52425 J\"ulich, Germany
}%
\author{Yunchao Qu}
\affiliation{
School of Traffic and Transportation, Beijing Jiaotong University, 100044 Beijing, China
}%
\author{Antoine Tordeux}
\affiliation{
University of Wuppertal, 42329 Wuppertal,Germany
}%
\author{Ziyou Gao}
\email{zygao@bjtu.edu.cn}
\affiliation{
School of Traffic and Transportation, Beijing Jiaotong University, 100044 Beijing, China
}%


\date{\today}

\begin{abstract}
In a crowd, individuals make different motion choices such as ``moving to destination", ``following another pedestrian", and ``making a detour". For the sake of convenience, the three direction choices are respectively called destination direction, following direction and detour direction in this paper. Here, it is found that the featured direction choices could be inspired by the shape characteristics of Voronoi diagram.  To be specific, in the Voronoi cell of a pedestrian, the direction to a Voronoi node is regarded as a potential ``detour'' direction, and the direction perpendicular to a Voronoi link is regarded as a potential ``following'' direction. A pedestrian generally owns several alternative Voronoi nodes and Voronoi links in a Voronoi cell, and the optimal detour and following direction are determined by considering related factors such as deviation. Plus the destination direction which is directly pointing to the destination, the three basic direction choices are defined in a Voronoi cell. In order to evaluate the Voronoi diagram based basic directions, the empirical trajectory data in both uni- and bi-directional flow experiments are extracted. A time series method considering the step frequency is used to reduce the original trajectories' swaying phenomena which might disturb the recognition of actual forward direction. The deviations between the empirical velocity direction and the basic directions are investigated, and each velocity direction is classified into a basic direction or regarded as an inexplicable direction according to the deviations. The analysis results show that each basic direction could be a potential direction choice for a pedestrian. The combination of the three basic directions could cover most empirical velocity direction choices in both uni- and bi-directional flow experiments.

\end{abstract}

\maketitle


\section{\label{sec:Section1} Introduction}
With the growing frequency of crowd activities, the investigation of pedestrian dynamics is attracting more attention. Understanding the principles of pedestrian motion is beneficial to the planning and designing of public facilities, as well as the schedule and organization of pedestrian crowds. Data analysis and simulation \cite{Schadschneider2009} are important methods for the understanding of pedestrian dynamics, and among them, the simulation method is considered to have many advantages such as low cost and high safety. 

Typical crowd motion modeling methods include force-based models \cite{Helbing1995,Chraibi2010} and cellular automata models \cite{Blue2001,Nowak2012}. In the force-based models, pedestrians are regarded as physical particles, and the Newton’s second law is applied. The sum of driving force and interaction force lead to the reaction of pedestrian motion. In cellular automata models, space is divided into discrete cells, and each cell would be occupied by one or several pedestrians. Based on specific rules, the pedestrian jumps from one cell to another to simulate the motion of crowd dynamics. 

Another modeling approach is the cognitive behavior method \cite{Gerd1999,Antonini2006,Michael2016}. The idea of this approach is to recognize the potential motion strategies in velocity choices and formulate the reasonable motion heuristics. A pedestrian normally has several different motion strategies such as maintaining the current velocity or making a turn. In \cite{Antonini2006}, a discrete choice framework for pedestrian walking behaviors is proposed. The velocity direction choice has been divided into several radial cones according to the deviation. Three speed choices, i.e., ``keep the same speed'',`` slow down'', and ``acceleration'' are proposed. In \cite{Michael2016}, the motion strategies are classified into four kinds, step or wait heuristic, tangential evasion heuristic, sideways evasion heuristic, and follower heuristic. The pedestrian will follow different heuristics depending on exact situations. On one hand, an advantage of these models is the convenience of considering the intelligent behaviors of a pedestrian. On the other hand, a core problem for this approach is the definition and cognition of the different pedestrian behaviors. 

Voronoi diagram \cite{Voronoi1908,Franz1991} is considered to have potential in both understanding pedestrian behavior \cite{Steffen2010,Zhang2011} and modeling pedestrian dynamics \cite{Xiao2016}. The Voronoi diagram is a partitioning of a plane into regions based on distance to points in a specific subset of the plane. Each region contains all the points closer to the related particle than to others. Due to its special geometric features, the Voronoi diagram has been applied in many fields, e.g., networking \cite{Ivan2006,Aziz} and biology \cite{Martin2010}. Also, motion planning is also a wide-used area for Voronoi diagram. In autonomous robot navigation \cite{Liao2003,Bhat2008,Santiago2006}, the Voronoi diagram is used to find feasible routes among obstacles. In pedestrian crowd experiments \cite{Zhang2011,Steffen2010,Ezaki2016}, the geometry features of Voronoi cell is used to calculate the local density of pedestrian. The method is capable of obtaining a fundamental diagram with fewer fluctuations. Concerning the modeling of pedestrian dynamics \cite{Xiao2016}, the Voronoi diagram of pedestrians is also introduced. According to three characteristic directions, i.e., destination direction, detour direction, and fine-tuning direction, a pedestrian recognition process is formulated for pedestrian motion. The simulation results show good agreement with the empirical fundamental diagram. 

In this work, based on the features of Voronoi diagram, three basic direction choices are introduced and defined. It is noted that the three basic direction choices proposed in this paper are a little bit different from the formal definitions \cite{Xiao2016}, mainly the following direction takes place of the fine-tuning direction. The combination of destination direction, following direction and detour direction, is believed to have a better performance in realizing the pedestrian behaviors. To investigate the effects of the basic directions, the empirical trajectory data in both uni- and bi-directional flow experiments are introduced and smoothed. Analyses from different aspects are presented in the text to explore the potential of the single basic direction, as well as the combination of the basic directions. 

The rest of the paper is organized as follows. Section 2 presents the definitions of the basic directions and an assessment method based on empirical data. In section 3, the setting of the trajectory experiments is introduced, and a smooth method is used to reduce the swaying in original trajectories. In section 4, the effects of the basic directions are investigated with the smoothed trajectories. Section 5 gives the conclusion and the prospect. 

\section{\label{sec:Section2} Basic direction choice and assessment method}
\begin{figure}
\includegraphics[width=0.5\textwidth]{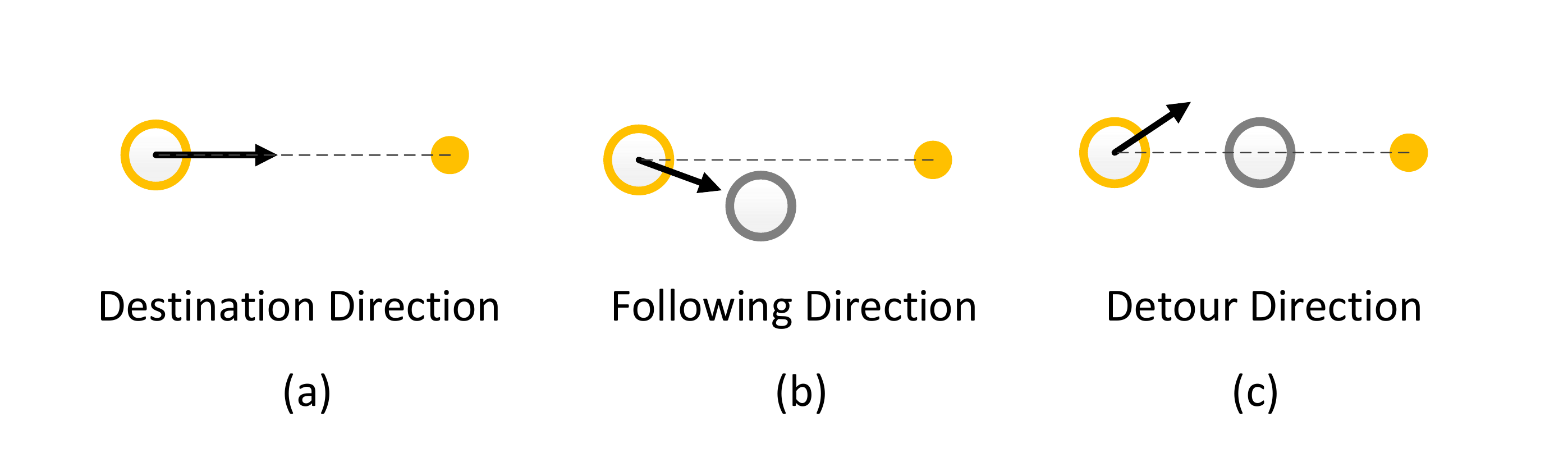}
\caption{\label{figdirectionchoices} Illustration of pedestrian motion direction choices. The dashed circles represent the pedestrian, and the solid dots represent the target of pedestrian.}
\end{figure}

In pedestrian crowds, three kinds of directions choices could be observed, which are ``moving to the destination'' ``following another pedestrian'' and ``making a detour''. For the sake of convenience, the three direction choices are called ‘ destination direction’, ‘ following direction’ and ‘ detour direction’, respectively (Fig.\ref{figdirectionchoices}). 

A pedestrian normally strives for a most efficient route to the destination, and the shortest route in geometry usually corresponds to the most efficient route. Since the destination direction points to the destination and indicates the shortest route, it is defined to be the default direction choice for motion \cite{Batty1997,Alasdair2002,Warren2009}. The following direction indicates the behavior along a neighboring pedestrian who has a similar motion pattern. The following direction might deviate from the shortest route, but most conflicts and collisions will be undertaken by the leader pedestrian while the follower is going to own a more comfortable walking environment. The detour direction generally deviates from the shortest route and points to an intermediate area between neighboring pedestrians. Usually, it is regarded as a regular option for the avoidance of conflicting/congestion area and the achievement of an overall efficient route. 

\subsection{\label{sec:section2.1} Voronoi diagram based directions}

\begin{figure}
\includegraphics[width=0.4\textwidth]{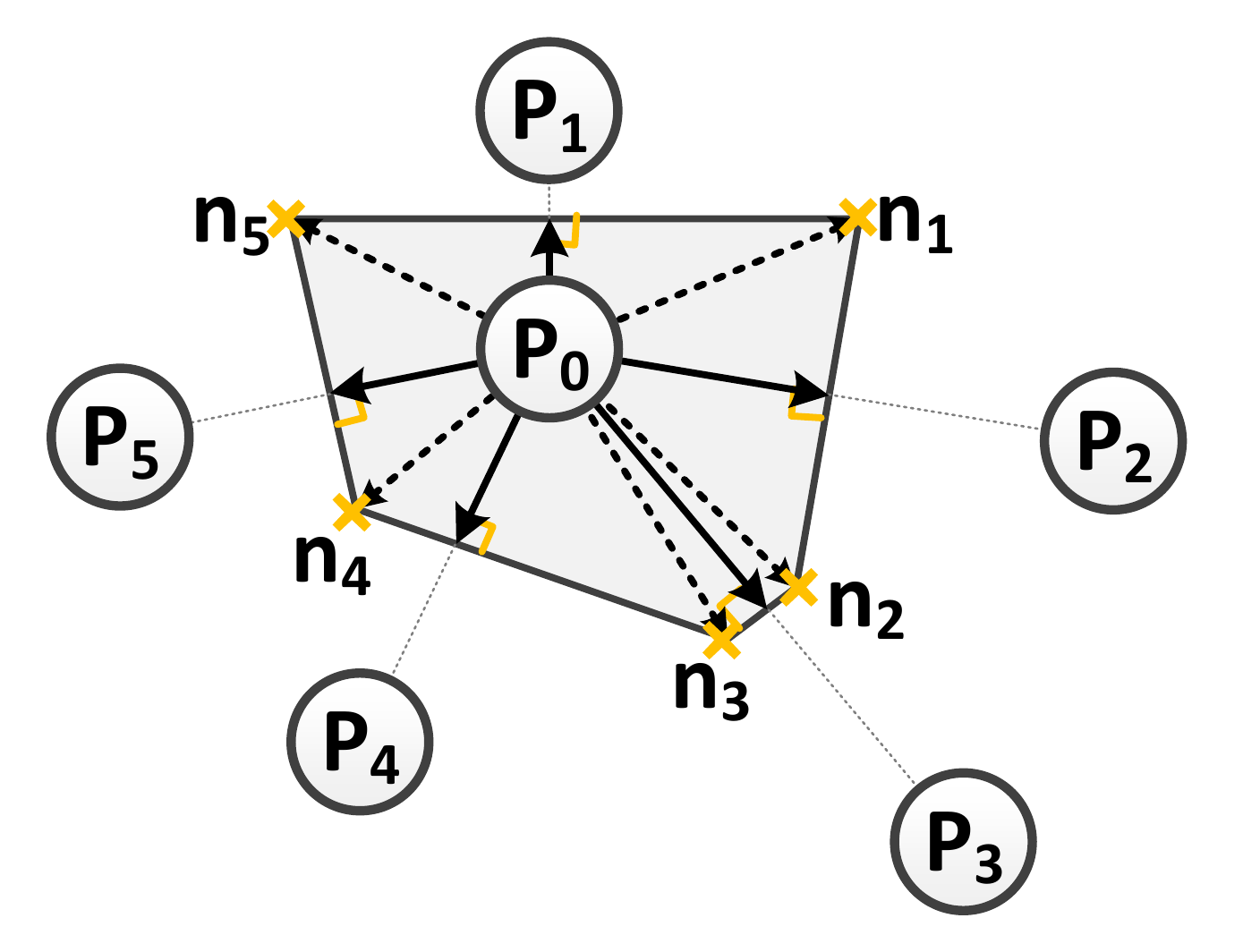}
\caption{\label{figvoronoidirection1} Potential following directions and potential detour directions.}
\end{figure}

The shape characteristics of Voronoi diagram inspires two kinds of basic direction choices for a pedestrian. Fig.\ref{figvoronoidirection1} shows a Voronoi cell of pedestrians to indicate the two kinds of directions. First, the direction pointing to the Voronoi node (dashed arrow) corresponds to the intermediate space between two neighboring pedestrians. Thus, it is defined as a potential detour direction. In this case, the pedestrian $P_{0}$ has five potential detour directions which are pointing to the five Voronoi nodes, $n_{1}$, $n_{2}$, $n_{3}$, $n_{4}$, $n_{5}$, respectively. Second, the direction perpendicular to the Voronoi links (solid arrow) corresponds to the neighboring pedestrians. Thus, it is defined as a potential following direction. In this case, the pedestrian $P_0$ has five potential following directions which are perpendicular to the Voronoi links. It is noted that these directions also point to the five neighboring pedestrians $P_{1}$, $P_{2}$, $P_{3}$, $P_{4}$, $P_{5}$ according to the properties of Voronoi diagram.

In this section, an evaluation function with respect to the pedestrian states is proposed to determine the optimal following and detour direction. The optimal following target is determined by, $P_{\rm{following}}^{*}=\underset{P_{j}\in M_{i}}{\mathrm{\argmax}}((\vec{e}_{i}\cdot \vec{e}_{j})\times(\vec{e}_{i}\cdot \vec{e}_{ij}))$, where $M_i$ represents the set of neighbors of pedestrian $P_i$, $\vec{e}_i$ and $\vec{e}_j$ are the unit vector of velocity of pedestrian $P_i$ and its following target $P_j$, respectively. $\vec{e}_{ij}$ is the unit vector of the direction from pedestrian $P_i$ to pedestrian $P_j$. Similarly, the optimal detour objective is determined by, $n_{\rm{detour}}^* = \underset{n_{j}\in N_{i}}{\mathrm{\argmax}}(\vec{e}_i\cdot \vec{e}_{ij}/\rho^n_j)$, where $N_i$ represents the set of the Voronoi nodes of pedestrian $P_i$. $\rho^n_j$ is the local density of Voronoi node $n_j$. In this paper, the local density of a Voronoi node is defined as the average value of the densities of its related pedestrians, and the detailed definition of local density could be found in Appendix.\ref{Appendixa}.

\begin{figure}
\includegraphics[width=0.5\textwidth]{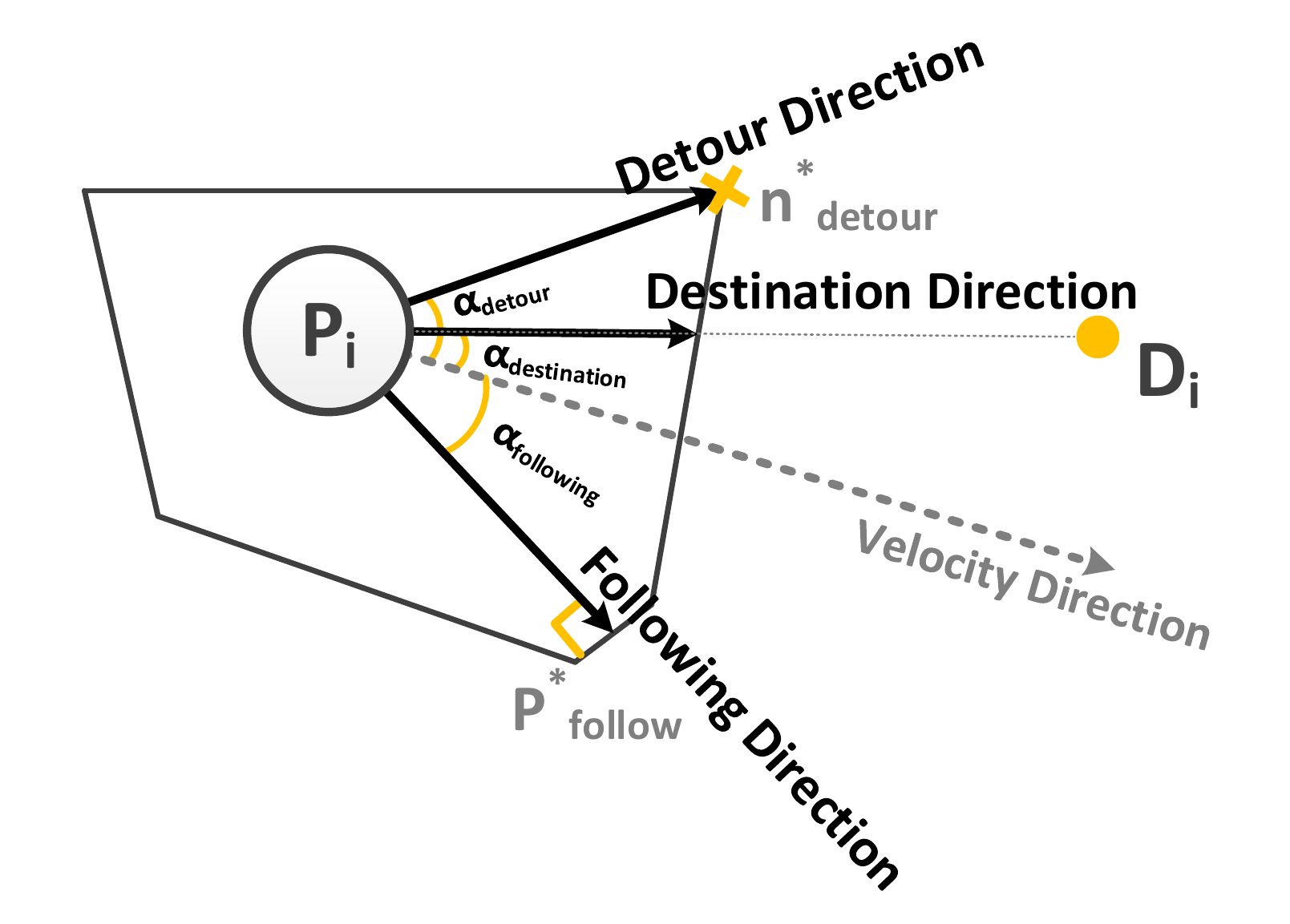}
\caption{\label{figvoronoidirection2} Voronoi diagram based direction choices.}
\end{figure}

The determination of destination may be a difficult problem in pedestrian simulation \cite{Antonini2006}. In some cases, the destination is changing with time and events. For example, in a shopping mall, the destination is easy to make a change along with the newfound attractors. In some cases, a pedestrian might lack a specific destination. For instance, some pedestrians might lose mind about the destination in an emergent evacuation situation. In this work, the destination is known and set up at first. As a result, the three basic direction choices, i.e. destination direction, following direction, and detour direction, are indicated in Fig.\ref{figvoronoidirection2}. The destination of pedestrian $P_i$ points to the target $D_i$, so the destination direction for pedestrian $P_i$ is given as, $\vec{e}_{\rm{dest}}^*=\overrightarrow{P_iD_i}/\|\overrightarrow{P_iD_i}\|$. The optimal following objective is obtained as $P_{\rm{follow}}^*$ with the method introduced in the last part, and the following direction is given as $\vec{e}_{\rm{follow}}^*=\overrightarrow{P_iP}_{\rm{follow}}^*/\|\overrightarrow{P_iP}_{\rm{follow}}^*\|$. The optimal detour Voronoi node is obtained as $n_{\rm{detour}}^*$ with the method just introduced, and the detour direction is given as $\vec{e}_{\rm{detour}}^*=\overrightarrow{P_in}_{\rm{detour}}^*/\|\overrightarrow{P_in}_{\rm{detour}}^*\|$.

\subsection{\label{sec:section2.2} Direction judgment method}
A direction judgment method is proposed in this section to classify the empirical velocity data. The empirical velocity direction data is determined as a basic direction or an inexplicable direction based on the deviation between them. The deviation from the velocity direction to the destination direction, following direction and detour direction, are calculated and called as destination deviation, $\alpha_{\rm{destination}}$, following deviation, $\alpha_{\rm{following}}$ and detour deviation $\alpha_{\rm{detour}}$, respectively (Fig.\ref{figvoronoidirection2}). $\alpha_{\rm{0}}$ is a threshold value for the judgment of a basic direction, and $\alpha_{\rm{min}}=\rm{min}⁡(\alpha_{\rm{destination}},\alpha_{\rm{following}},\alpha_{\rm{detour}},\alpha_{\rm{0}})$. As a result, the classification of the velocity direction is given as, 

$$C=\left\{
\begin{array}{rcl}
\rm{Destination} \quad \rm{direction},     &      & {\alpha_{\rm{min}}=\alpha_{\rm{destination}}}\\
\rm{Following} \quad \rm{direction},       &      & {\alpha_{\rm{min}}=\alpha_{\rm{following}}}\\
\rm{Detour} \quad \rm{direction},	        &      & {\alpha_{\rm{min}}=\alpha_{\rm{detour}}}\\
\rm{Inexplicable} \quad \rm{direction},    &      & {\alpha_{\rm{min}}= \alpha_{\rm{0}}}\\
\end{array}.\right.$$

Note that $\alpha_{\rm{0}}$ is critical for the classification, and it should not be too large or too small. With a strict value, for instance, $\alpha_{\rm{0}}= \pi/180 $, very few empirical data could be classified as a basic direction. With a loose value, for instance, $\alpha_{\rm{0}}=\pi/3$, some deflecting data might be classified as basic directions. A more detailed analysis could be found in Appendix.\ref{Appendixb}, and in this case, $\alpha_{\rm{0}}$ is set to $\pi/18$. 

\section{\label{sec:section3}Experimental data}

In this section, the data from uni- and bi-directional flow experiments \cite{Zhang2011,Zhang2012} were used for the investigation of Voronoi diagram based direction choices. 

\begin{figure}
\includegraphics[width=\linewidth]{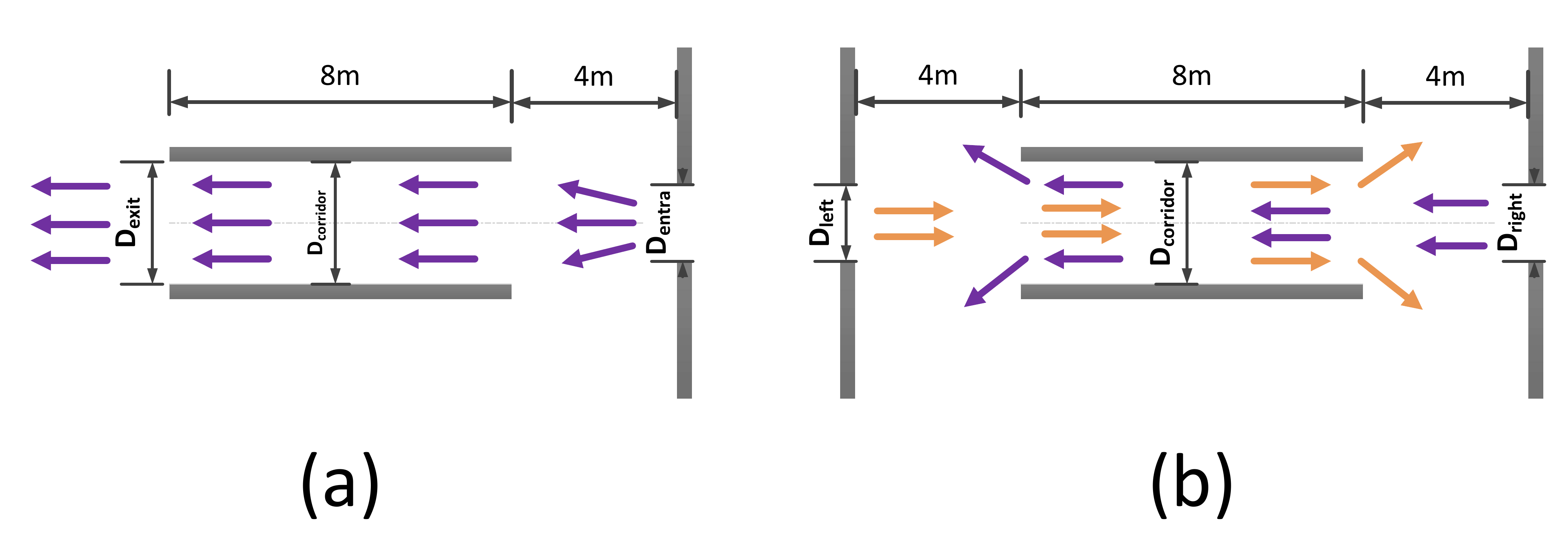}
\caption{\label{geometries} (a)Sketch map of the uni-directional flow experiment. (b)Sketch map of the bi-directional flow experiment.}
\end{figure}

The geometry configurations of the uni- and bi-directional flow experiments are shown in Fig.\ref{geometries}. In the uni-directional flow experiments (Fig.\ref{geometries}(a)), the length of corridor is 8m (constant), and the width of the corridor, i.e., $D_{\rm{corridor}}$, is changing among 1.8m, 2.4m and 3m. In these experiments, the width of the exit $D_{\rm{exit}}$ is equal to the width of corridor $D_{\rm{corridor}}$, so there is no bottleneck in the corridor and the destination of a pedestrian is easy to be obtained. According to the specific features of corridor experiments, the destination direction is defined to be parallel to the wall of corridor and points to the opposite end of the corridor. At the initial stage, the pedestrians are waiting outside the entrance. Through the adjustment of entrance width $D_{\rm{entra}}$, the uni-directional pedestrian flow into the corridor can be controlled. In the bi-directional flow experiments (Fig.\ref{geometries}(b)), the length of corridor is 8m (constant), and there are two values for the corridor width $D_{\rm{corridor}}$ which are 3m and 3.6m, respectively. The pedestrians are waiting outside the entrance at the two sides, and the bi-directional pedestrian flow into the corridor is able to be controlled by the width of entrance $D_{\rm{left}}$ and $D_{\rm{right}}$. It is noted that there is a buffer zone (length = 4m) between the entrance and the corridor in both uni- and bi-directional flow experiments. The buffer zone is used to minimize the effect of entrance bottleneck so that the pedestrians could be homogeneously distributed on the total width of the corridor. 

\begin{figure}
\includegraphics[width=0.5\textwidth]{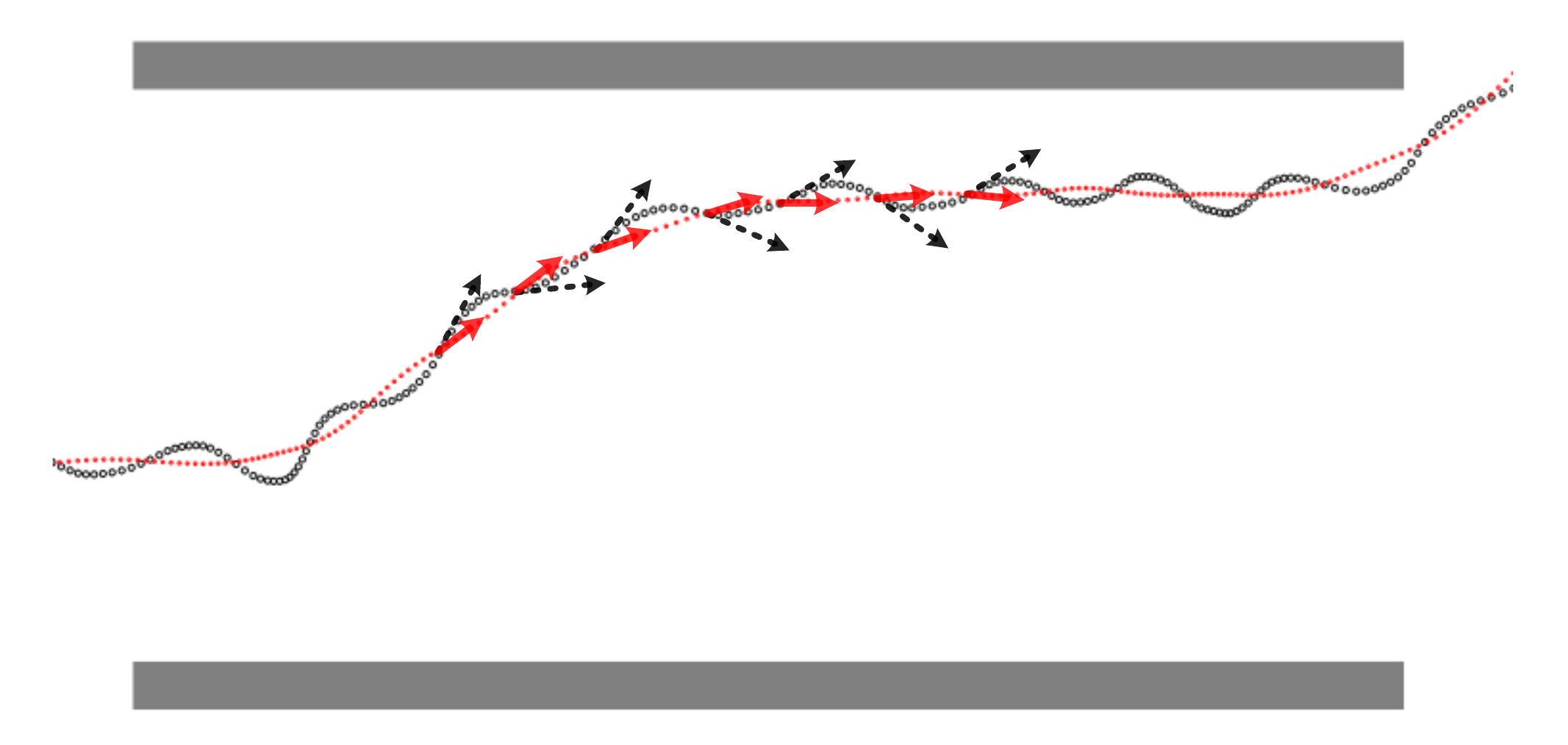}
\caption{\label{trajectorysmoothness} An example of the original trajectories and the smoothed trajectories (color). The black points represent the original trajectory data and the red points represent the smooth trajectory data. The red and black arrows represent the direction based on the original and smoothed trajectories.}
\end{figure}

In the experiments, the crowd motions are recorded by the cameras mounted on the ceiling, and the location of the head of a pedestrian is obtained at each frame step. An example of the original trajectories is shown as the black points in Fig.\ref{trajectorysmoothness}. The trajectories pattern is similar to an oscillation curve, and the zigzag feature in the trajectories is usually called swaying phenomenon \cite{Hoogendoorn2005,Kim2004,Grieve1966}. The phenomenon is caused by the step walking behaviors of the pedestrian. A pedestrian naturally needs to shift the body to maintain balance during walking. Thus, the body especially the head has to sway left and right with different foot striking the ground, and it leads to the zigzag trajectories. Here, the velocity direction is determined as the direction from its current location to the location of the next time step, and the original trajectories based velocity directions (dashed arrow) fluctuate frequently and highly in Fig.\ref{trajectorysmoothness}. As a result, due to the swaying phenomenon, the real forward direction of a pedestrian is likely to be covered in the original trajectories and it might terribly affect the judgment of the direction choice classification. To exclude the influence of swaying phenomenon, a time series method is introduced. The original location of the pedestrian at time step $t$ is assumed as $\vec{l_t}=(x_t, y_t)$, and the smoothed location is given as, 
$$\vec{l_t'}=(\sum_{i=t-k}^{t+k}x_i/(2k+1), \sum_{i=t-k}^{t+k}y_i/(2k+1)),$$
where $k$ is a free parameter, and it is found work best when corresponds to the step length (Appendix.\ref{Appendixb}). To efficiently reduce the swaying effect, the step frequency is considered and $k$ is given by, $k=f_c/f_s+1$. Where $f_c$ is the frames per second of the camera, and $f_c$ = 16 (1/s) in our uni- and bi-directional flow experiments. The step frequency $f_s$ is estimated by \cite{Grieve1966}, $f_s=1.72 \cdot bv^p/h $, where h is the height of pedestrian, $v$ is the value of current speed, $b$ and $p$ are two dimensionless parameters, $b$ = 1.57 and $p$ = 0.5. 

Based on the smoothed trajectories, the specific velocity direction (solid arrow) is determined as shown in Fig.\ref{trajectorysmoothness}. The fluctuations of the velocity direction caused by the swaying trajectories are reduced significantly. The set of velocity direction data in uni- and bi-directional flow experiments formulate the basis of analysis in the work. 

\section{\label{sec:section4}Results Analysis}

\begin{figure}
\includegraphics[width=0.5\textwidth]{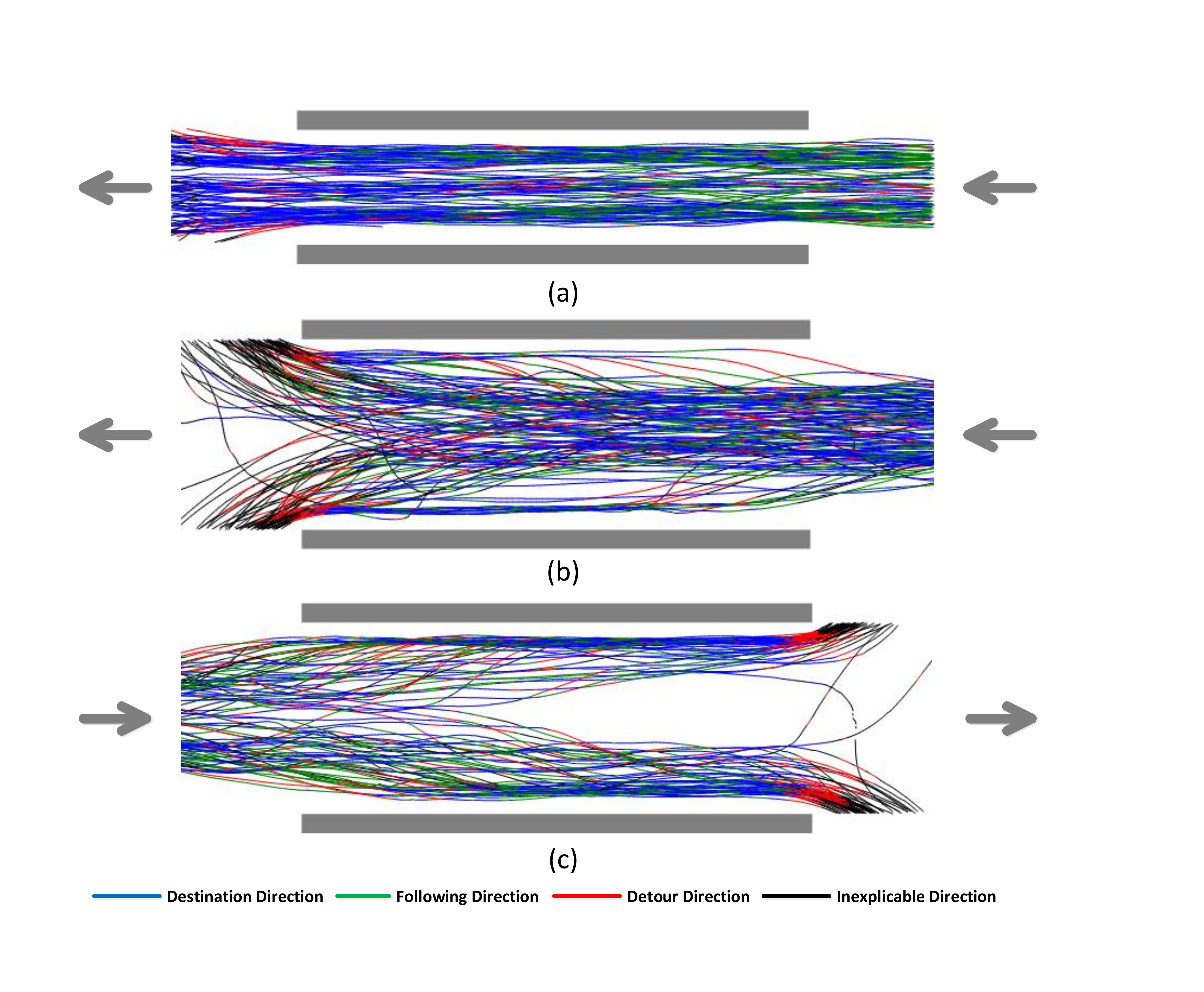}
\caption{\label{trajectories} Featured trajectories in uni and bi-directional flow (color). The trajectories are colored with blue, green, red and black to represent destination, following, detour and inexplicable direction, respectively. (a) Pedestrian trajectories of uni-directional flow experiment. The parameters of the uni-directional flow corridor parameters are, $D_{\rm{entra}} = 0.7\,\rm{m}$ and $D_{\rm{corridor}}=D_{\rm{exit}}=1.8\,\rm{m}$. (b) Pedestrian trajectories from right to left in bi-directional flow experiment. (c) Pedestrian trajectories from left to right in bi-directional flow experiment. The parameters of the bi-directional flow corridor are, $D_{\rm{corridor}} = 3.6\,\rm{m}$ and $D_{\rm{left}}=D_{\rm{right}}=0.5\,\rm{m}$.}
\end{figure}

In this section, the empirical pedestrian velocity data in both uni- and bi-directional flow experiments are used for the investigation of the Voronoi diagram based basic direction. According to the direction judgment method, the empirical direction data have been classified into four classes, the three basic directions (destination direction, following direction, detour direction) and the inexplicable direction. In Fig.\ref{trajectories}, the trajectories are shown with different colors to represent the direction classification, and the qualitative direction pattern in the corridor is found. Among them, the blue color represents the defined destination direction, the green color represents the defined following direction, the red color represents the defined detour direction and the black color represents the inexplicable direction. 

Fig.\ref{trajectories}(a) shows the featured pedestrian trajectories in uni-directional flow experiment. In the experiment, entrance width equals to 0.7m and 111 pedestrians enter the corridor from the right end to the left end. In general, the destination direction and following direction play the dominant roles, and the detour direction and inexplicable direction are limited to a small percentage. Both the destination direction and following direction are found at the entrance(right) side of the corridor, while the destination direction plays a much more significant role at the exit(left) side and very few directions are still obtained as the following direction. It is noted that the pedestrians have spread to a wide width at the buffer zone. Thus, the entrance(right) side of corridor plays as a kind of bottleneck for the pedestrians, and the pedestrians are likely to prefer the following behavior to enter the bottleneck. After the entering, both conflicts and obstacles are very rare in the corridor. The pedestrian just needs to move forward with the shortest route which corresponds to the destination direction, and the following behavior is not so required. As a result, the following directions are gradually reduced from the entrance side to the exit side. 

Fig.\ref{trajectories}(b) and Fig.\ref{trajectories}(c) show the featured pedestrian trajectories to the right side and the left side, respectively. The entrance width equals to 0.5 m and 130 pedestrians taking part in the bi-directional flow experiment. The pedestrian motion pattern to the left side and the right side are generally not identical. In this experiment, the pedestrian flow to the left side separates quite late in the corridor, while the pedestrian flow to the right side separates earlier. However, the separation pattern is not so stable and it may vary among different experiments. In addition, the constitution of basic directions in the bi-directional flow experiment is more complicated than the uni-directional flow experiment, but the destination direction and following direction also play important roles especially in the center part of the corridor. At the corner of the corridor, the inexplicable direction and detour direction rises due to a special requirement of the experiment. The pedestrians are required to leave the corridor from the corner, and the required direction is usually deviating from the default destination which is parallel to the wall in our method. To maintain the simplicity of the destination setting and the validity of the data, the trajectory data at the end of the corridor are removed from our analyses. 

\begin{figure}
\includegraphics[width=0.5\textwidth]{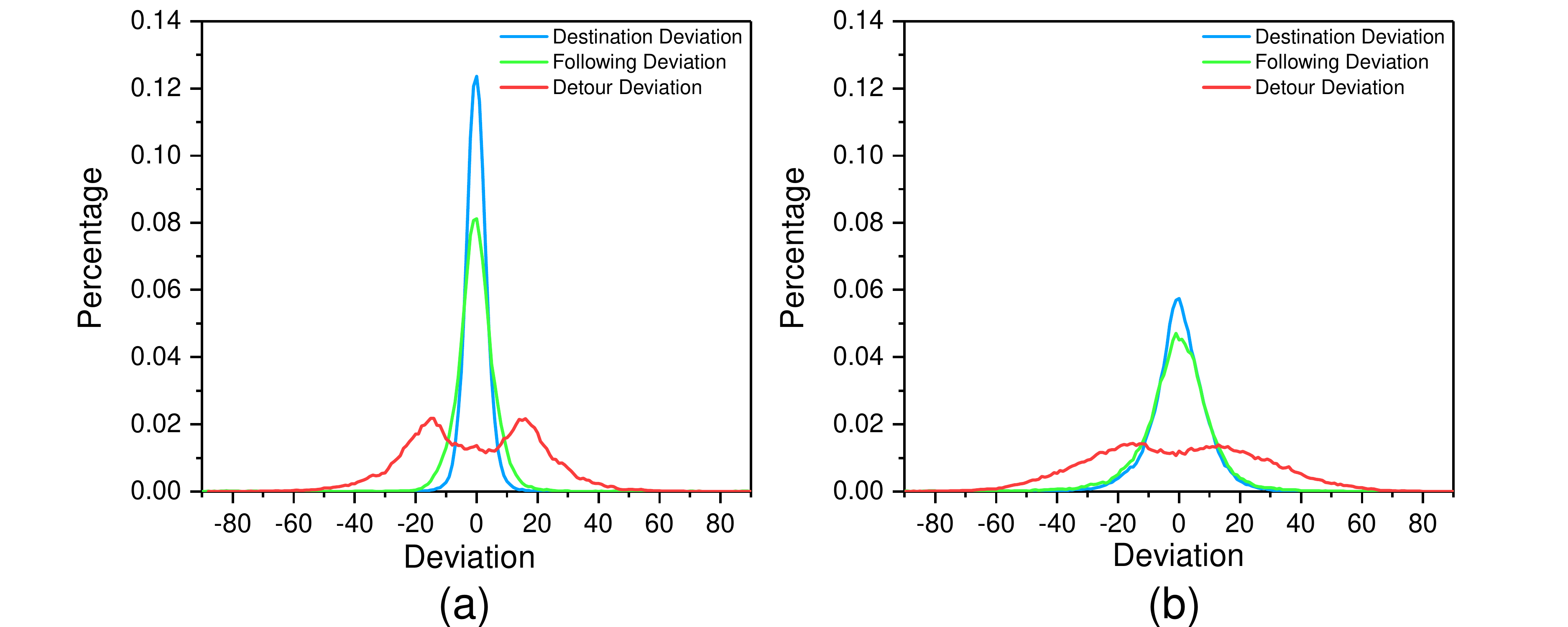}
\caption{\label{deviation} Distribution of deviation between current velocity direction and basic directions. (a) unidirectional flow experiment data (b)bidirectional flow experiment data.}
\end{figure}

\begin{figure}
\includegraphics[width=\linewidth]{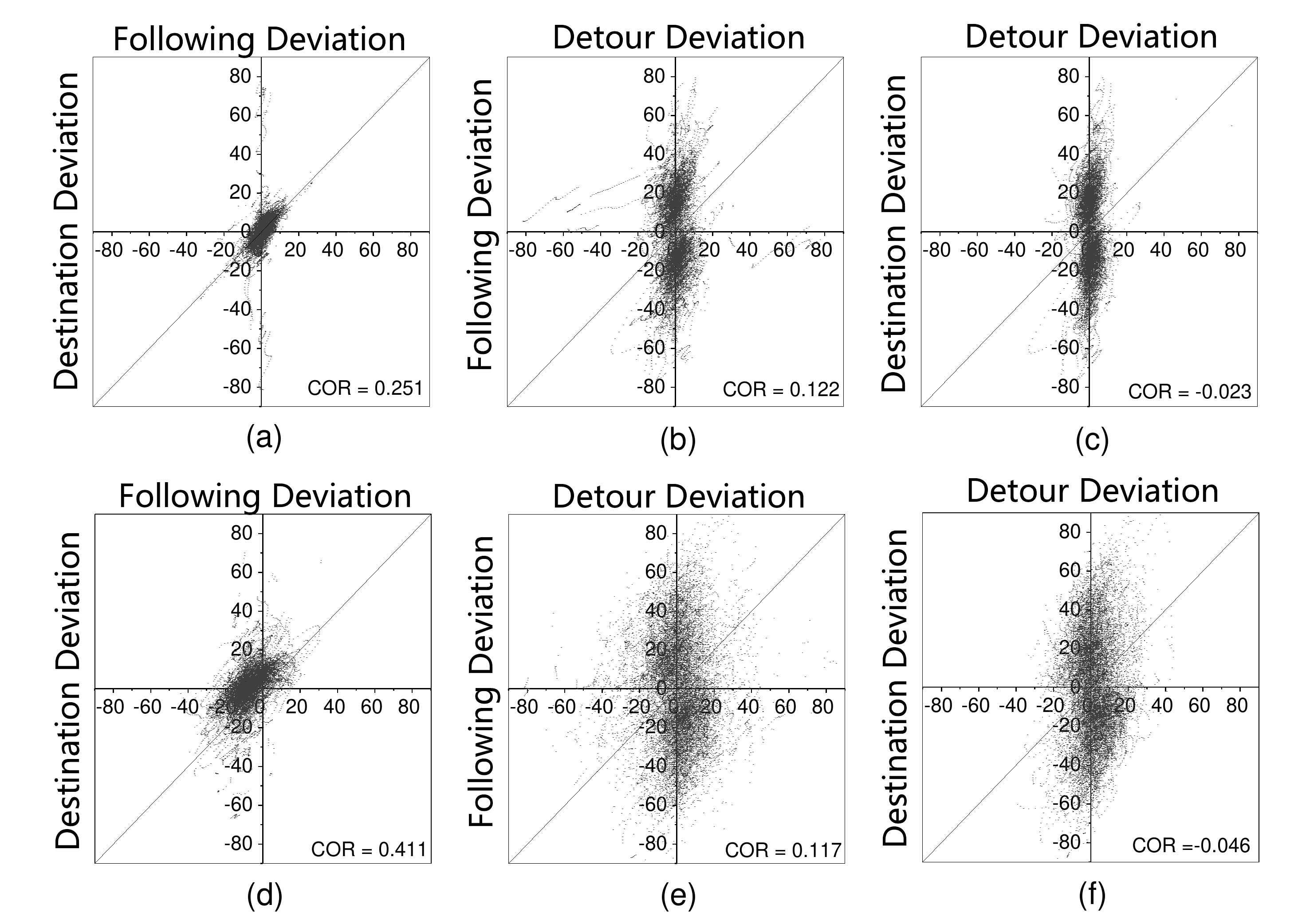}
\caption{\label{correlation} Correlation of deviation between basic directions. (a) and (d) are destination deviation-following deviation relation in uni- and bi-directional flow experiments, respectively. (b) and (e) are following deviation-detour deviation in uni- and bi-directional flow experiments, respectively. (c) and (f) are destination deviation-detour deviation in uni- and bi-directional flow experiments, respectively.}
\end{figure}

\begin{figure*}
\includegraphics[width=\textwidth]{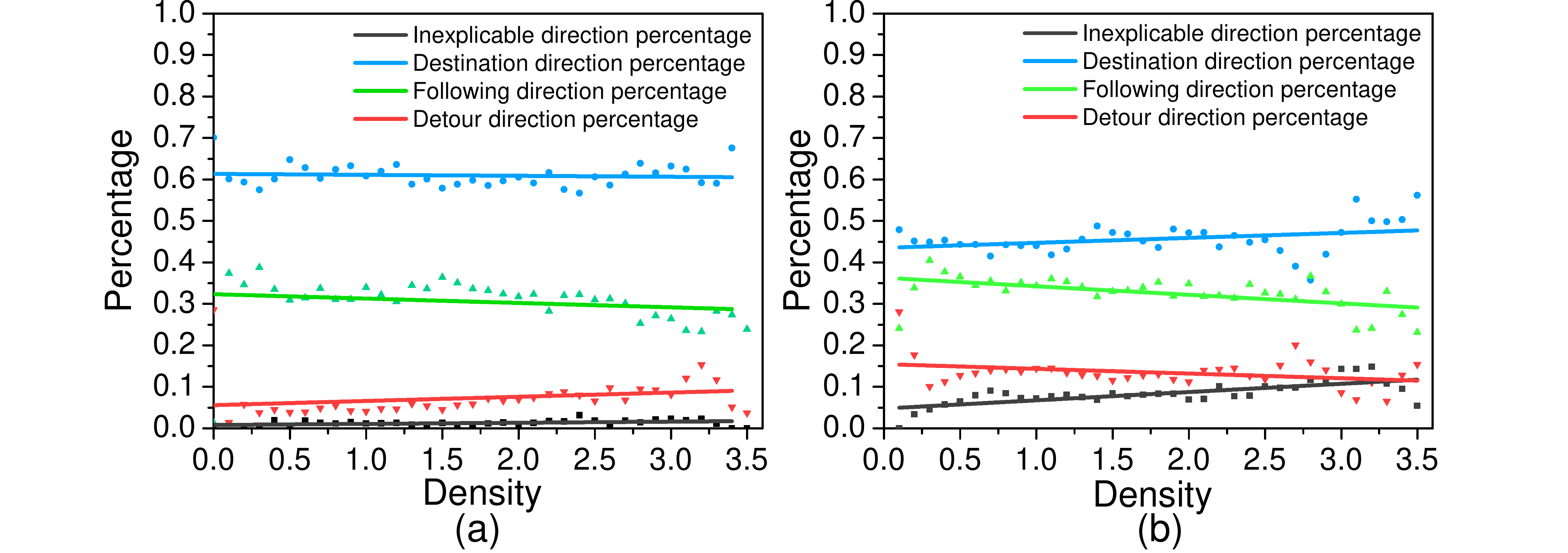}
\caption{\label{percentage} Percentage of basic directions in different densities (a) Basic direction percentage for uni-directional flow. (b) Basic direction percentage for bi-directional flow. Note that the density of pedestrian is obtained by a Voronoi method which is able to calculate the local density of single pedestrian, the detail definition could be found in Appendix\ref{Appendixa}.}
\end{figure*}

For the quantitative features of the three basic directions, base data contains 19 uni-directional flow experiments and 18 bi-directional flow experiments which own different corridor width or entrance width \cite{Zhang2011,Zhang2012}. Angular deviation is the most direct quantitative way to measure the differences between directions. Therefore, the angular deviation between defined basic directions and empirical direction data are first explored. In Fig.\ref{deviation}, the deviation distribution of the three basic directions in uni- and bi-directional flow experiments are obtained. In general, all kinds of distributions in both experiments are symmetrical and range between around -60 to 60 degrees. In the uni-directional flow experiments, the distribution of destination deviation is centered over 0 degree and spreads from about -20 to 20 degrees in the uni-directional flow experiments. The pedestrian rarely deviates a lot from the destination direction since almost no serious conflicts and obstacles exist in the uni-directional flow. Although overtaking the slow pedestrian might take place in the uni-directional flow, the pedestrian prefers to choose a soft and gradual way to achieve it. In bi-directional flow experiments, the distribution of destination deviation is also centered over 0 degree, but the distribution spreads in a wider range, about -40 to 40 degrees. It is known that the conflicts are more common and frequent in the bi-directional flow, especially that the pedestrian has to face the oncoming pedestrian from the other side. Hence, the pedestrian has to make urgent and sharp adjustments to the current velocity to avoid these conflicts. The central tendency of destination deviation in the experiments implies that the destination direction is a significant direction choice. The distributions of the following deviation are similar to the distributions of destination deviation, which are also centered over 0 degree in both uni- and bi-directional flow experiments. The following direction is also likely to be an important direction choice. The distributions of detour deviation get two peak values at about -15 and 15 degrees in both uni-directional flow and bi-directional flow. According to the characteristics of Voronoi based detour direction, its deviation to the destination direction is likely to be around 15 degrees. Meanwhile, the destination direction and following direction are the main choices by most pedestrians most of the time in corridor experiments. In conclusion, the choices for the frequently used basic directions lead to the peak values in the distribution of the detour deviation.

\begin{table}[ht]
\caption{\label{tab1}Correlation coefficient between basic directions}
\centering
\begin{tabular}{|l|c|c|c|}
\hline
 	& Dest-Following  & Dest-Detour & Following-Detour \\
\hline
Uni- & 0.251 & 0.122 & -0.023 \\
\hline
Bi- & 0.411& 0.117 & -0.046  \\
\hline
\end{tabular}
\end{table}

As found in Fig.\ref{deviation}, the distributions of destination direction and following direction are quite similar. In order to investigate the correlation of basic directions, the deviation from the current velocity direction to the basic directions in uni- and bi-directional flow experiments are compared in Fig.\ref{correlation}. Here, the Pearson correlation coefficient is introduced and applied as, 
$COR_{X,Y} = \sum_{i=1}^n(x_i-\overline{x})(y_i-\overline{y})/\sqrt[2]{\sum_{i=1}^n(x_i-\overline{x})^2}\sqrt[2]{\sum_{i=1}^n(y_i-\overline{y})^2}$
where data set $X=(x_1,x_2,\cdots,x_n)$ and $Y=(y_1,y_2,\cdots,y_n)$ are the deviation data set of a basic direction, respectively. $\overline{x}$ and $\overline{y}$ are the mean deviation of the data set $X$ and $Y$. First, the absolute value of correlation coefficient between destination deviation and detour deviation, as well as the correlation coefficient between following deviation and detour deviation are less than 0.15. It indicates that these two sets of deviations of basic directions are not strongly correlated in both uni- and bi-directional flow experiments. The absolute value of correlation coefficient between destination direction and following direction, especially in bi-directional flow, is remarkably larger than the other two kinds of coefficients. The specific value in bi-directional flow could be explained by the lane formation phenomenon. In bi-directional flow, most pedestrians would walk within a lane to obtain a comfortable walking environment, and the destination direction naturally agrees with the following direction in the case. Therefore, there is an idea that these two directions might have no difference in the case. To test it, the null hypothesis $H_0$ is given as that the destination direction and following direction in bi-directional flow have no difference. A paired t-test is presented here, and the significance level is set as 0.05 and two-tailed. The calculation results of 108164 pairs data show that $t = 16.781 > t_{critcal} = 1.959$, that is to say, the p-value is almost zero. Therefore, the null hypothesis $H_0$ is rejected. The conclusion should be the alternative hypothesis $H_1$ that these two kinds of direction choices have some differences in bi-directional flow.

Next, the performance of the combination of basic directions is investigated, and the percentage of different basic directions are given as Fig.\ref{percentage}. In uni-directional flow (Fig.\ref{percentage}(a)), more than 95\% direction data are explained as the three basic directions. Among them, the destination direction plays a dominant role (around 60\%). Since the motion conflicts rarely occur in uni-directional flow, pedestrians are not likely to deviate from the shortest route. Following and detour direction take up around 30\% and 7\% of the total data, respectively. With increasing densities, it is observed that the following percentage tends to decrease while the detour percentage tends to increase. The reduction of the personal space leads to the rising desire to obtain a more comfortable personal space, so the pedestrian is more likely to make a detour to change the current position instead of following in a more crowded situation. 

In bi-directional flow (Fig.\ref{percentage}(b)), around 90\% percent direction data are explained as the three basic directions. The destination direction also plays a most important role (around 45\%) in the total direction data, but the percentage is less dominating compared with the uni-directional flow data. The main reason is that the conflicts are much more frequent in the bi-directional flow compared with the uni-directional flow experiment. Thus, the detour direction grows to be a more important choice for directly dealing with the frequent conflicts. At the same time, adopting the following direction is another effective method to avoid the conflicts, and that is a critical reason for the appearance of lane formation phenomenon in the bi-directional flow. As a result, the percentage of detour direction and following direction increase in the bi-directional flow. There are two reasons for the growing of inexplicable direction percentage in the bi-directional flow. First, the pedestrian dynamics are richer in the bi-directional flow. Second, the pedestrians are required to leave the corridor from a specific corner in the bi-directional flow experiment, so the actual destination direction might differ from the original destination direction which is parallel to the wall. Adjusting the definition of destination in corridor might reduce the inexplicable percentage, but maintaining a simple definition of the destination is also useful. With increasing densities, the following direction percentage decreases and the inexplicable direction percentage increases. Similar to the uni-directional flow, the reduction of personal space is considered to be the reason for this change. 

\section{\label{sec:section5}Conclusion and prospect}

Three kinds of direction choices are found to describe the different pedestrian motion patterns. They are destination direction, i.e., the direction moving to the destination, following direction, i.e., the direction following another pedestrian, and detour direction, i.e., the direction for making a detour. The three basic directions are considered to be important direction choices in the crowd motion. Inspired by the characteristics of Voronoi diagram of pedestrians, the three kinds of basic directions are determined for each pedestrian at each time step. The empirical trajectory data in uni- and bi-directional flow experiments are used here to investigate the effects of the three directions. A time series method is introduced to smooth the original trajectories and obtain a reasonable forward velocity. Based on the smoothed trajectory data, the velocity direction is determined and classified into a specific basic direction whenever possible. 

Based on the smoothed trajectories in uni- and bi-directional flow experiments, the features of the three basic directions are qualitatively and quantitatively investigated, including the direction pattern, the deviation distribution, the correlation between basic directions, and the percentage of different basic direction choices. It is found that the direction choice patterns in the corridor experiments are able to be recognized by the basic directions. First, different proportions of basic directions are obtained in different parts of the corridor. Second, the motion direction patterns are significantly different between uni- and bi-directional flow experiments. 

The results also show that the general velocity directions could be reduced into the three kinds of basic direction choices. In addition to the three basic directions, more direction choices are still possible. For example, the detour direction is able to be classified into two categories, one for the avoidance of local collisions and another for the planning of a global optimal route. The two kinds of detour direction might point to a similar direction, but the motivations vary. Another problem for our work is the simplicity of corridor experiments. With only two kinds of destinations in the corridor, the crowd motion patterns could be quite simple. For instance, the destination and following directions are quite similar in these cases. Although these two kinds of choices are proved to be different, it raises an idea that the two kinds of directions might be merged in pedestrian motion. To completely deal with the problem, further analyses in even more complicated scenarios, e.g., bottleneck\cite{Iker2014,Takahiro2012} and evacuation situation\cite{Guo2012,Tang2015}, are scheduled for the next-step investigation of the basic directions. A further concern is regarding the pedestrian behaviors in different density situations. A pedestrian is able to make intelligent motion choice in a low-density environment, whereas it would be much more difficult for the pedestrian to determine the movement in a crowded situation since the passive collision force between the pedestrians might dominate the motion. In this case, these defined direction choices might lose their effectiveness. These different combinations of basic direction choices need further empirical investigations and analyses. Besides, simulations based on the basic direction choices could be presented to explore the potentials of the different combinations, and it would be our next-step work. 

The introduction of the three basic directions could inspire the modeling of pedestrian dynamics. An important problem for the pedestrian dynamics is to determine the velocity at the next time step, and the velocity determination could be divided into two procedures, velocity direction determination and speed determination \cite{Antoine2016,Moussaid2011}. The three basic directions are inspired by the pedestrian cognitive heuristics, and they have the potential to summarize the general pedestrian direction choices. Based on it, determining the direction choice based on the cognitive process of a pedestrian is a promising modeling method. In \cite{Michael2016}, four kinds of heuristics, i.e., step or wait heuristics, tangential evasion heuristic, sideways evasion heuristic, and follower heuristic are proposed to represent the different pedestrian behaviors. Similarly, a simple logit based model could be given based on the three basic direction choices. Many factors such as velocity, density and deviation could be considered into the utility function and support the choice of the logit method.   

In conclusion, the investigation with Voronoi diagram based directions is a meaningful work for the pedestrian motion research. The understanding of pedestrian motion is promoted by the Voronoi diagram based directions. Also, it inspires the work for lots of fields such as the modeling of pedestrian dynamics. 

\section*{Acknowledgment}

The study is supported by the Foundation for Innovative Research Groups of the National Natural Science Foundation of China (71621001), the National Natural Science Foundation of China (71601017) and the Chinese Postdoctoral Science Foundation (2017M610042). In addition, we wish to thank Nikolai Bode, Arne Graf, Erik Andresen and Armin Seyfried for the useful discussions and valuable suggestions.

\begin{appendix}
\section{\label{Appendixa}Local density}
\begin{figure}[!ht]
\includegraphics[width=0.3\textwidth]{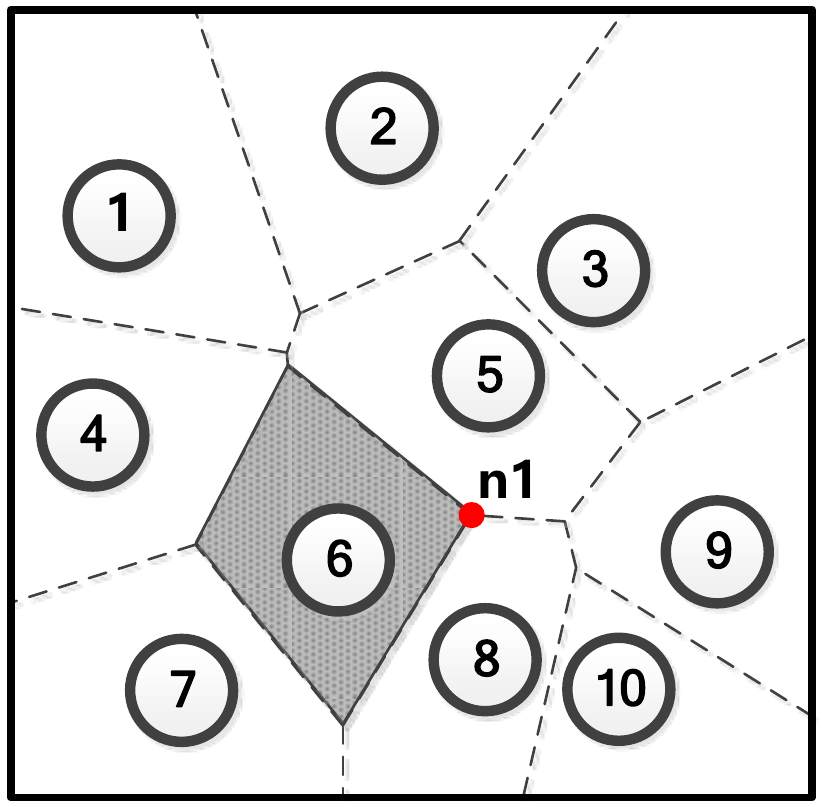}
\caption{\label{figlocaldensity} Voronoi diagram of pedestrians.}
\end{figure}

This section contains the definitions of local density for both pedestrian and Voronoi node in the Voronoi diagram of pedestrians. In Fig.\ref{figlocaldensity}, a Voronoi diagram of pedestrians is shown. There are ten pedestrians represented by the dashed circles, and the Voronoi diagram is drawn based on them. The area of the Voronoi cell of pedestrian $P_i$ is given as $a_i$. For instance, the Voronoi cell area of pedestrian $P_6$ is the shadow area which equals to $a_6$. In our method, the local density of pedestrian $P_6$ is defined as the reciprocal of the area of its corresponding Voronoi cell, $\rho^p_6=1/a_6$. It makes sense since the Voronoi cell actually contains all the closest space of the pedestrian. Moreover, the local density of Voronoi node is defined as the average density of its related pedestrians, $\rho^n_j= \sum_{i=1}^m(\rho^p_i)/m$. Note that the related pedestrians are those pedestrians who own this Voronoi node. For Voronoi node $n_1$(Fig\ref{figlocaldensity}), its local density $\rho^n_1 = (\rho^p_5 +\rho^p_6 +\rho^p_8)/3$.

\section{\label{Appendixb}Sensitivity Analysis}
\begin{figure}[!ht]
\includegraphics[width=0.5\textwidth]{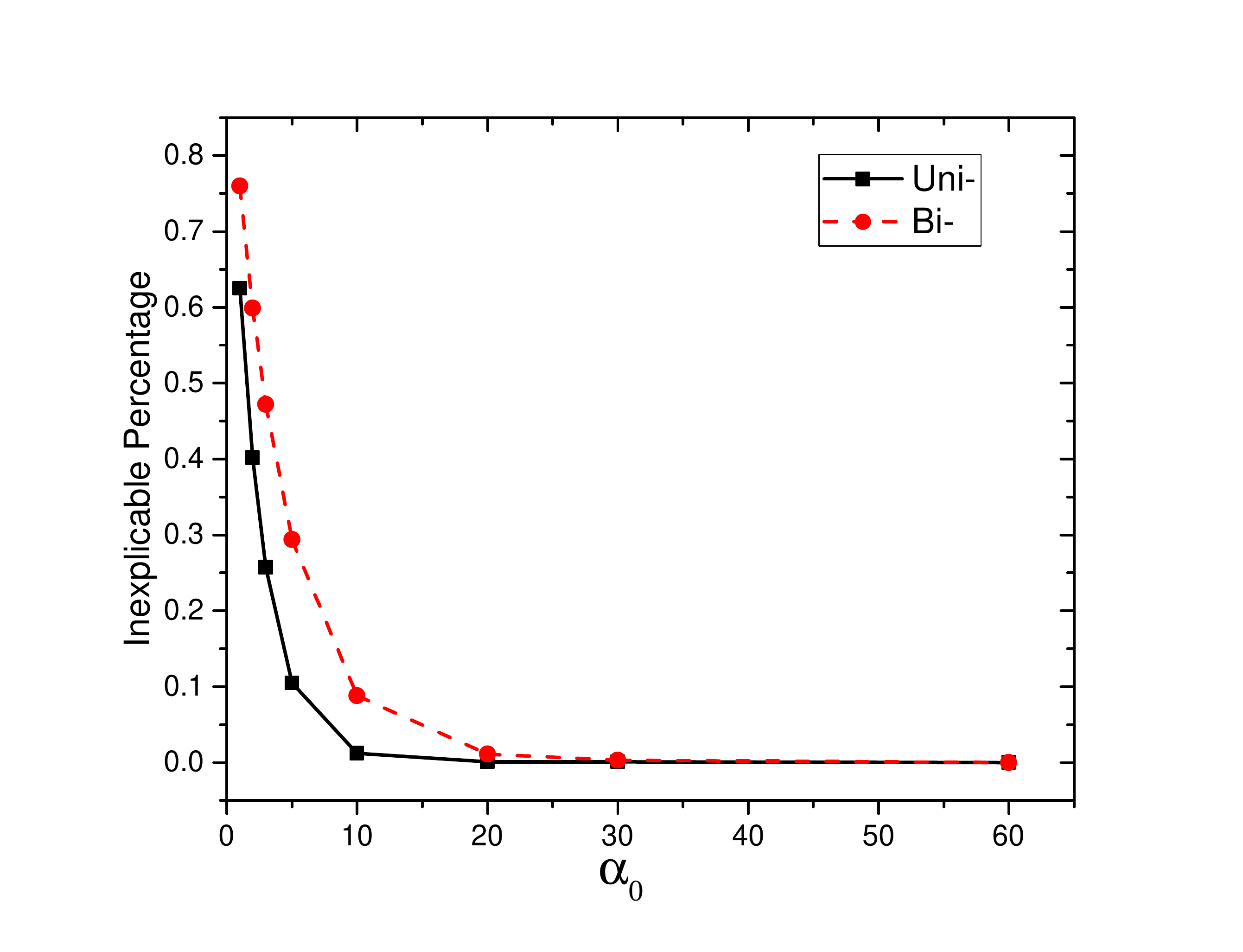}
\caption{\label{figAlpha} Relationship between $\alpha_0$ and percentage of inexplicable direction. Note that $\alpha_0$ is measured in degrees in the figure.}
\end{figure}

This section contains the sensitivity analyses of both $\alpha_0$ and $k$. The threshold value $\alpha_0$ is critical for the classifications of basic directions. Fig.\ref{figAlpha} shows the relationship between $\alpha_0$ and the percentage of inexplicable direction in basic direction classifications for both uni- and bi-directional flow experiments. In both experiments, the percentages of inexplicable direction decrease with the growing of $\alpha_0$. Apparently, $\alpha_0$ cannot be limited to be too strict. A mass of empirical trajectory data might be ignored due to the swaying phenomenon and other errors. In the case of $\alpha_0=\pi/180=1\ \rm{degree}$, only 20-30 percent trajectory data are able to be distinguished as basic directions. Also, $\alpha_0$ cannot be too loose. In the case of $\alpha_0=\pi/3=60\ \rm{degree}$, those deflecting trajectory data might be classified as basic directions, and almost 100 percent empirical trajectory data are explained. Therefore, based on the limitations of both sides, $\alpha_0$ is set to $\pi/18$ in this paper.

\begin{figure}[!ht]
\includegraphics[width=0.5\textwidth]{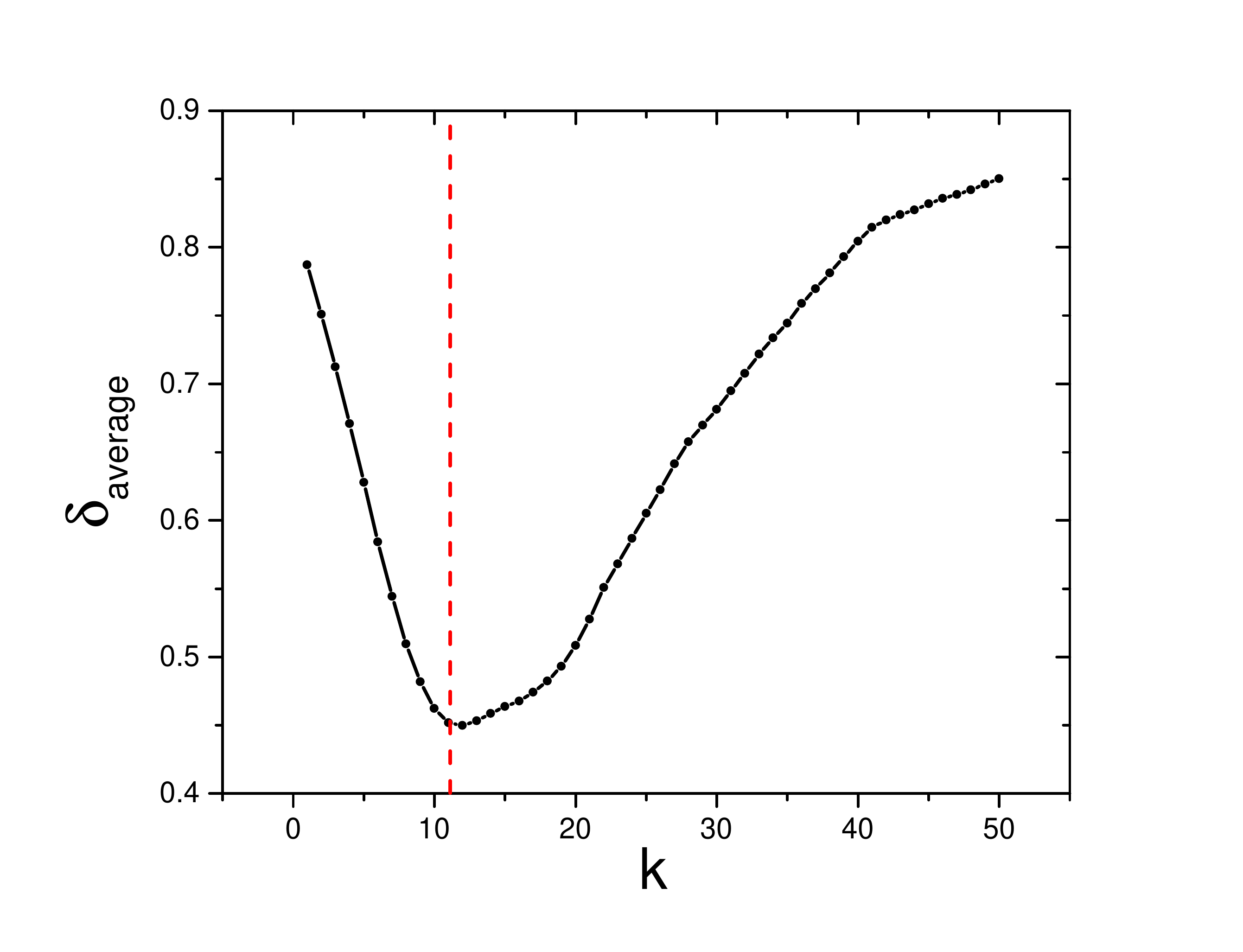}
\caption{\label{figSmooth} Relationship between smoothing steps and average direction deviation. The average step length is marked by the red dashed line}
\end{figure}

Qualitatively, the shape of the original pedestrian trajectories is similar to a sine function due to the swaying phenomenon (see Fig.\ref{trajectorysmoothness}). The real motion direction is difficult to be obtained based on the oscillating trajectories, so a time-averaging method is proposed for the smoothness. $k$ is a core parameter in the time-averaging method, and it represents the smooth time step. In order to achieve the smoothness of trajectories and keep the necessary velocity tendency, $k$ is best to correspond to the period of trajectories for the smoothness. To perform the sensitivity analysis of $k$, a parameter $\delta_i$ is introduced. $\delta_i$ is the maximum angular deviation between current velocity direction and its adjacent velocity direction data on the set of trajectories of a pedestrian, 
$$\delta_i=\{\max \limits_{x}\langle \vec{v}_x, \vec{v}_i \rangle, i-j\leq x \leq i+j, x\in \mathbb{N}\}.$$
Where the adjacent velocity data parameter $j=25$ in this section. $\delta_i$ could be used to investigate the deviation level of trajectories, in other words, the smoothing effect. $\delta_{average}$ is the average value of $\delta_i$ for the whole trajectory set, $\delta_{average}=\sum_{i=1}^n(\delta_i)/n$. $\delta_{average}$ could be used to measure the effect of different smooth parameters. As shown in Fig.\ref{figSmooth}, $\delta_{average}$ firstly decreases with the growing of smooth steps and reaches a minimum value around $k=12$. It's found that $k=12$ basically corresponds to the calculated average step length (Fig.\ref{figSmooth}) \cite{Grieve1966}. As a result, match the smoothing steps to the step length is likely to smooth the pedestrian trajectories.

\end{appendix} 
\bibliography{main}

\end{document}